\documentclass[11pt,twoside]{article} 
\usepackage{asp2004}
\usepackage{epsf}
\usepackage{psfig}
\usepackage{lscape} % for landscape figures/tables

\begin{document} 
\title{Subluminous O stars from the ESO Supernova Progenitor Survey - Observation versus Theory}
\author{A. Stroeer$^1$, U. Heber$^1$, 
  T. Lisker$^{1,3}$, R. Napiwotzki$^{1,4}$ and S. Dreizler$^2$} 
\affil{$^1$Dr. Remeis-Sternwarte Bamberg, Astronomical Institute of
    the University of Erlangen-N\"urnberg, Sternwartstra\ss e 7,
    D-96049 Bamberg, Germany\\
$^2$University of G\"ottingen, Geismarlandstr. 11, 37083 G\"ottingen, Germany\\
$^3$Institute of Astronomy, ETH
    Z\"urich, Departement of Physics, HPF D8, ETH H\"onggerberg, 8093
    Z\"urich, Switzerland\\
$^4$Departement of
    Physics and Astronomy, University of Leicester, University Road,
    Leicester LE1 7RH, UK}
% please use mixed case type and Initials-Familyname order
% The last author's name must be proceeded by 'and'.
% If you have many authors and institutions you may want to use
% superscripts instead and collect all institutions at the end of the
% author list. In this case the superscript should follow the
% separating comma 
% \author{A. Author1,$^1$ B. Author2,$^2$ C. Author3,$^3$ 
% and D. Author4$^3$}
% \affil{$^1$Institute1 with Postal Address\\
% $^2$Institute2 with Postal Address\\
% $^3$Institute3 with Postal Address\\}
\begin{abstract} 
The ESO Supernova Progenitor Survey (SPY) has identified 58
(mostly helium-rich) subluminous O stars. We use the Balmer line strength 
to distinguish sdO from He-sdO (no Balmer lines) and present the
results of the analyses of high resolution optical VLT-UVES
spectra using an extensive grid of NLTE atmosphere models 
covering a large range in T$_\mathrm{eff}$, $\log{g}$ and helium 
abundances. The stellar atmospheric parameters are derived from line profile 
fits using a $\chi^2$ technique. The resulting distribution in the 
(T$_\mathrm{eff}$, $\log{g}$) diagram as well as the luminosity function are 
discussed in the context of stellar evolution scenarios. By combining our 
results with those for the sdB stars from SPY (Lisker et al.~2004) we discuss 
the implications for binary population synthesis models of Han et al.~(2003).
Models with a low CEE efficiency and a constant mass ratio distribution 
provide a reasonable explanation of the observed properties 
of the SPY sample of sdB and sdO stars indicating that the sdO stars
form the hot and luminous extension of the sdB sequence. 
However, for the He-sdO stars none of the considered evolution 
scenarios are in agreement with the measured parameters of our programme 
stars. We conclude that He-sdO stars are formed by a different process than 
the sdB and sdO stars.

\end{abstract}
\section{Introduction}
The ESO Supernova Progenitor Survey (SPY; Napiwotzki et al.~2003) has 
identified 137 hot
subluminous stars. 79 of them were classified as hydrogen rich sdB stars, 
58 of them were classified as subluminous O stars.
The sdB spectra have been studied by Lisker et al.~(2004, see also
these proceedings). 
%A new extensive grid of NLTE models was
%calculated to perform spectral analyses. 
%Introducing new results from the
%ESO SPY project, Lisker et al. (2004) discussed the distribution of sdB stars
They tested several evolutionary scenarios by comparing the results of their 
spectroscopic analyses
to model predictions. In particular, the 
binary population synthesis models of Han et al.~(2003, hereafter HPMM)
were discussed. However, two diagnostic tools used yielded conflicting 
results. 
%It turned out, that 
%the cumulative luminosity function does
%not match with the theoretically predicted one from Han et al. (2002, 2003) 
%at the luminous end. 
Moreover, the HPMM models predict that some EHB stars 
should be even hotter than those found in the sdB star sample of
Lisker et al.~(2004). This led the authors to the conclusion that the subdwarf
sample may not be sufficiently complete to describe the whole parameter range 
covered by the
simulations. 
By including the subluminous O
stars we attempt here to correct for this observational bias in the study of 
Lisker et al.~(2004).

\section{Spectral classification and analysis}\label{analy}
High resolution spectra covering the spectral range from 3300\,\AA \ to
6650\,\AA \ at a resolution of 0.36\,\AA \ have been obtained with the UVES
spectrograph at the ESO-VLT. 
In the course of spectral classification we distinguished 
He-sdO (30) from sdO stars (28) by the absence of Balmer
line absorption in the former.

An extensive grid of NLTE atmosphere models was calculated using the
latest version of the PRO2 code (Werner \& Dreizler 1999) that employs a
new temperature correction technique (Dreizler, 2003). A new detailed 
model atom for helium appropriate for the sdO temperature regime was
constructed.
%{\bf Stefan, koenntest du da etwas genaueres dazu angeben, evtl. Zahl der
%NLTE-Niveaus, Linienuebergaenge{\ldots} )}.  
2700 partially line blanketed NLTE model atmospheres consisting of hydrogen and helium 
were calculated 
resulting in
a grid of unprecedented coverage and resolution, extending from 
%resulting in a grid that extends from 
30\,000~K to
100\,000~K in T$_\mathrm{eff}$, from 4.0 to 6.4 in $\log{g}$ and 
from $-$4 to 3 in
helium abundance $\log{N_{He}/N_H}$ in order to
match the diversity of observed spectra. The step sizes are 2\,000~K from
30\,000~K to 52\,000~K and 5\,000~K from 55\,000~K to
100\,000~K; 0.2 and $\sim$0.5 dex, respectively.

Effective Temperatures (T$_\mathrm{eff}$), surface gravities ($\log{g}$), and 
helium
abundances ($N_{He}/N_H$) for 49 stars were determined by fitting 
simultaneously
hydrogen and helium lines to our synthetic model spectra, using a
$\chi^2$~procedure (Napiwotzki, 1999).
Resulting temperatures range from 36000~K to 78000~K, gravities from
$\log{g}$=4.9 to 6.4, and helium abundances from
$\log{N_{He}/N_H}$=-3 to +3. Four stars have temperatures in
excess of 60000~K and are probably post-AGB stars and will not be
discussed further.
%Since the intention of SPY is to search for radial velocity variable stars, 
%two or more exposures of each star were available. Thus the final fit
%results were calculated as the mean value of the results, weighted
%with the S/N level of individual exposures. The quality of most spectra is 
%excellent.
%the median S/N being about 100.
%The results together with the sdB stars from SPY analyzed by
%Lisker et al. (2004) are presented in Fig.~\ref{tefflogg} in the (T$_\mathrm{eff}$, $\log\left(g\right)$)-diagram.

\begin{figure}
\vspace{8.0cm}
\includegraphics{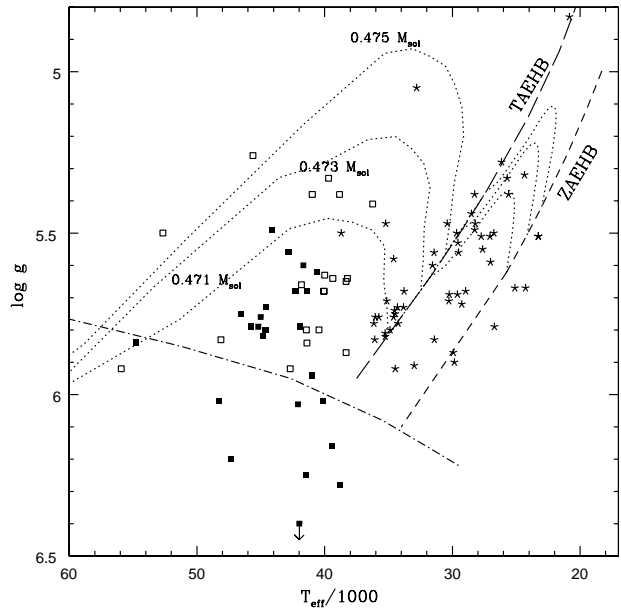}
\caption{Distribution of parameters of our analyses of sdO and He-sdO stars,
  compared to all sdB stars analyzed in Lisker et al.~(2004). 
  Filled squares show
  He-sdO stars, open squares sdO stars and asterisks represent sdB stars. 
  The location of the EHB band, the helium main sequence and
  tracks for post-EHB evolution (dotted; 
  Dorman et al. ~1993) 
  are also shown.}
\label{tefflogg}
\end{figure}

%\begin{figure}[ht!]
%\plotone{wd_as_disc_cI.eps}
%\caption{Distribution of parameters of our analyses of sdOs and HesdOs, also
%  compared to all sdBs analyzed in Lisker et al. (2004). Filled squares show
%  HesdOs, open squares sdOs and asterisks represent sdBs.}
%\label{tefflogg}
%\end{figure}

\section{Evolutionary status}\label{disc}
In Fig.~\ref{tefflogg} we compare 
the derived atmospheric parameter for the combined 
sample of subluminous O and B stars to the EHB band, 
the helium main sequence and canonical tracks for 
EHB and post-EHB evolution. 
%As can be seen, stars at higher temperatures but lower
%gravities  could be identified to be post-EHB
%stars. Thus we exclude those stars for the following discussion. 
%From Fig.~\ref{tefflogg} it is obvious,
%that HesdOs do display a different distribution than sdOs and sdBs. In
%addition the HesdOs show a different trend for the helium abundance with
%effective temperatures compared to sdB and sdO stars. 
%The cumulative luminosity function instead does
%show nearly the same shape for HesdOs and sdOs compared to sdBs  in the
%mid-region with a shift
%to higher luminosities. sdOs towards higher luminosities can
%be found to be in evolved stages, HesdOs towards lower luminosities to be
%positioned below the He-ZAMS,  resulting in a flatten of the function.
%Any evolutionary scenario has to explain not only the position of the star
%in the (T$_\mathrm{eff}$, $\log{g}$) diagram but also the helium
%enrichment.
%We tried to interpret the results for those sdO and HesdO stars in the
%context of 
%three evolutionary scenarios, i.e. (i) the late hot flasher 
%scenario (Sweigart 1997), (ii) the close
%binary scenario of Driebe et al. (1998) and (iii) 
We shall focus here, however, on
the binary population
synthesis models of HPMM and discuss other possibilities briefly in Sect. 3.2.

\subsection{Binary population synthesis}

\begin{figure}[ht!]
%\plotone{hpmm-vgl-a-I.eps}
\plotone{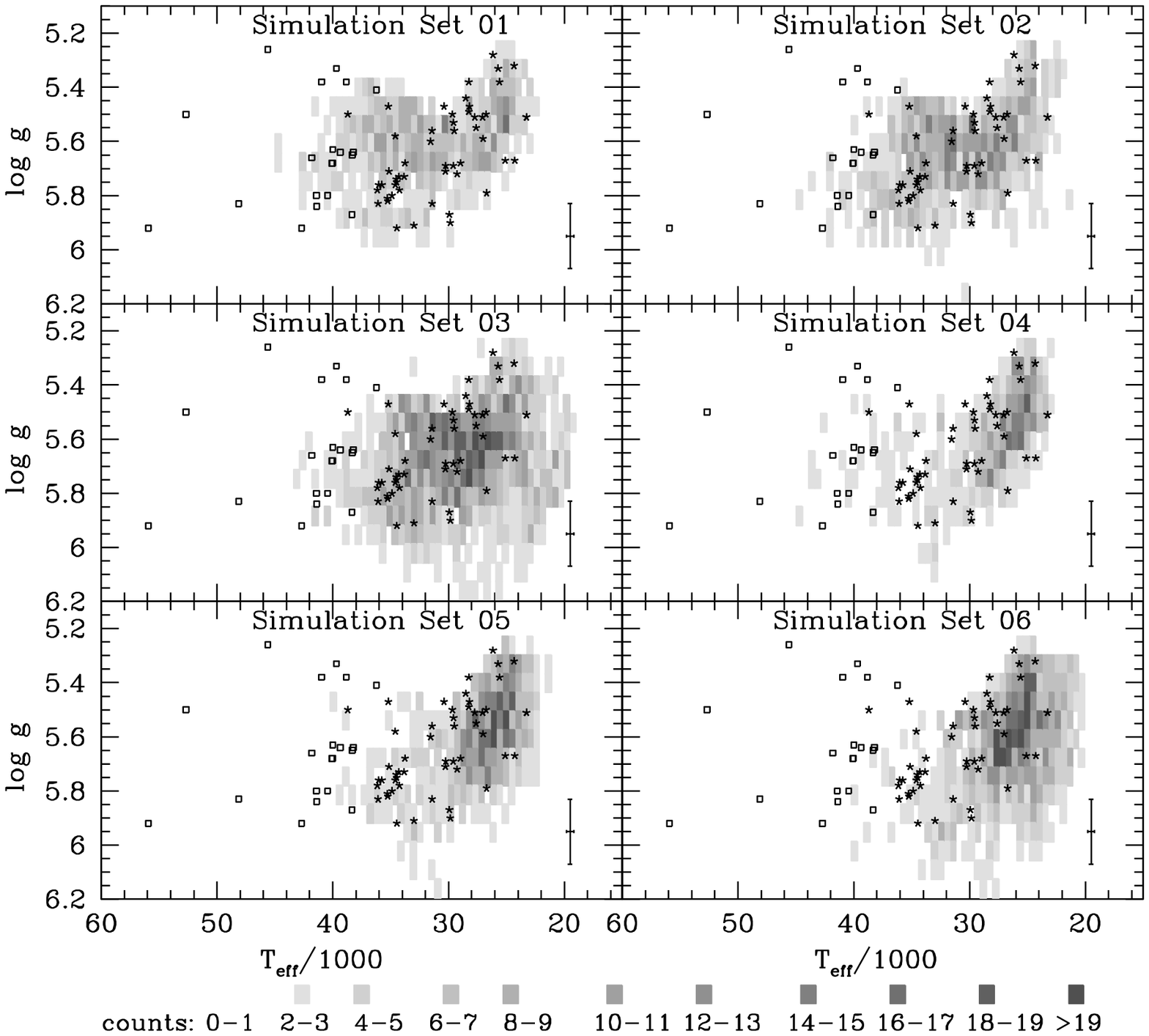}
\caption{Comparison of atmospheric parameters for sdO and sdB stars 
  with six of the
  HPMM  simulation sets. The theoretical predictions are shown as shaded boxes,
  where a higher density per box corresponds to darker shadings. In all sets
  solar metalicity and a critical mass ratio q for the first stable RLOF of
  q=1.5 are used. In sets 1{\ldots}3 a constant mass-ratio distribution 
  was adopted whereas in sets 4{\ldots}6 the component masses are 
  uncorrelated. The CEE efficiency is 50\% in sets 1\&4, 75\% in sets 2\&5
  and 100\% in sets 3\&6.   
}
\label{scen}
\end{figure}

HPMM calculated close binary population synthesis models and showed 
that core helium burning EHB stars can form via 
(a) stable Roche lobe overflow (RLOF), 
(b) common envelope ejection (CEE) and (c) by the merging of two He white 
dwarfs (see Lisker et al.~2004 for details). 
Since we do not know a priori whether the sdO and the He-sdO
stars belong to the same population or not we studied two hypotheses. 
%We tested the HPMM simulation sets assuming two
%different hypotheses. 
In hypothesis I, sdO and
He-sdO stars were treated as separate populations. 
Thus only sdO and sdB stars were 
considered to belong to the
EHB (see Fig.~\ref{scen}). 
Hypothesis II treats sdO and He-sdO stars as being of the same 
origin, thus also He-sdO stars were
considered to belong to the EHB. 

\begin{figure}
\vspace{6.5cm}
%\special{psfile=proceedcumlum2.eps hscale=34 vscale=34 hoffset=-5
%voffset=-55 angle=0}
%\special{psfile=proceedcumlum1.eps hscale=34 vscale=34 hoffset=+185
%voffset=-55 angle=0}
\includegraphics{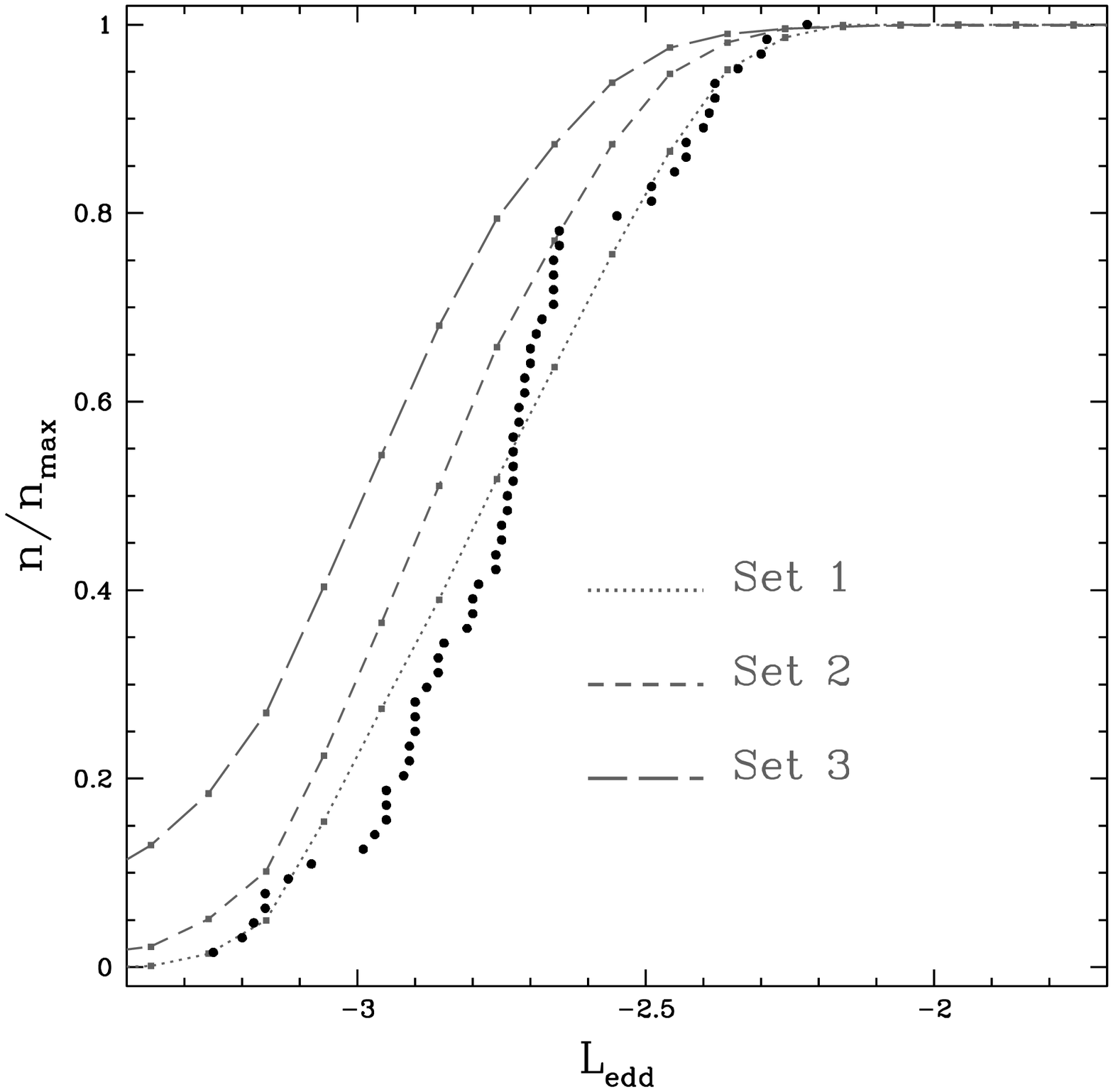}
\includegraphics{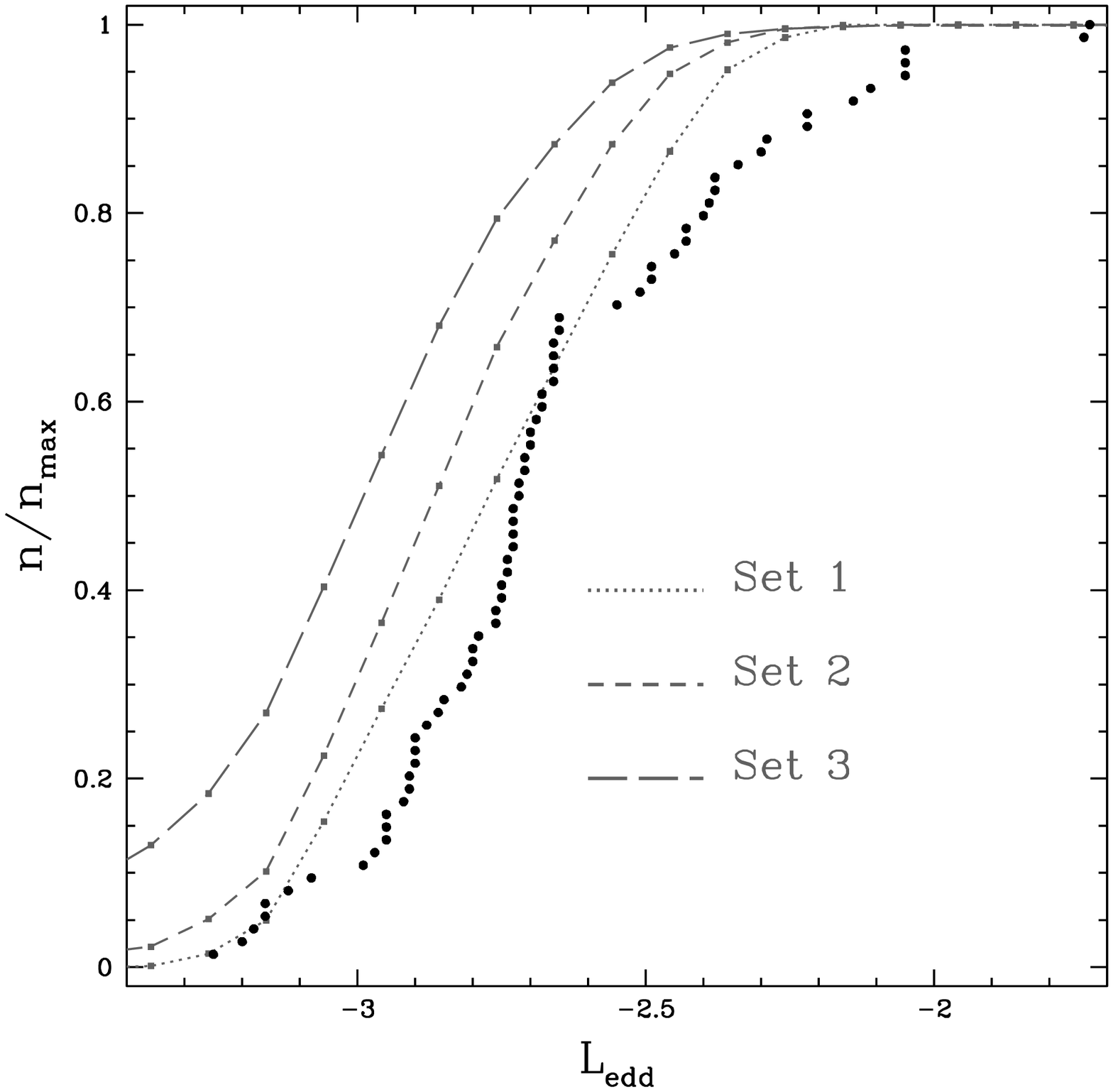}
\caption{Cumulative luminosity functions for the combined sdB plus sdO sample 
(hypothesis I) for three of the HPMM simulation sets. 
The luminosity is given in units 
of the Eddington luminosity. For the model parameters see Fig.~\ref{scen}.
{\bf Left} hand side: {\bf a)} all sdO stars included; {\bf right} hand side: 
{\bf b)} potential post EHB stars excluded.}
\label{cumlum}
\end{figure}

%\begin{figure}[ht!]
%\plotone{cumlumsdosdbcomp1.eps}
%\emph{Cumulative luminosity function}
%\end{figure}

We compared our results to all 12 simulation sets of HPMM 
%via visual 
%inspection 
in the (T$_\mathrm{eff}$, $\log{g}$)-diagram 
as well as by making use of the cumulative luminosity function.
%Fig. \ref{scen} illustrates this in the (T$_\mathrm{eff}$,
%$\log{g}$)-plane for six of the simulation sets and 
%Fig. \ref{cumlum} in the case the luminosity function for three simulation 
%sets. Details of the models are given in the caption of 
%Fig.~\ref{cumlum}.

\subsubsection{Hypothesis I: sdB and sdO stars combined.}

%Adopting hypothesis I (sdO + sdB) an improvement the comparison with all 
%simulation sets
%improved in comparison with the sdB-only sample (Lisker et al. 2004). 
%The 
%same holds for 
Fig.~\ref{scen} compares the sdB + sdO sample in the (T$_\mathrm{eff}$,
$\log{g}$)-plane to six of the HPMM simulation sets and 
Fig.~\ref{cumlum} its luminosity function to three simulation 
sets. Details of the models are given in the caption of 
Fig.~\ref{scen}.
Some of the HPMM models match the observed density distribution in the
(T$_\mathrm{eff}$, $\log{g}$) plane reasonably well (e.g. set 1).
The location of the four hottest sdO stars in the sample, however, cannot
be matched by any of the HPMM simulation sets. Another four sdO stars and 
one sdB
have gravities too low to be explained by any of HPPM's simulation sets.
These stars may have evolved beyond core helium burning (post-EHB, see also
Fig.~\ref{tefflogg}).
Based on evolutionary time scale estimates we would expect about 10\% of our
sample of 72 stars to be post-EHB stars, roughly consistent with
the number of stars (9) outside the predicted area in the (T$_\mathrm{eff}$,
$\log{g}$) diagram.
The cumulative luminosity function tends to be shifted to higher 
luminosities than predicted (see Fig.~\ref{cumlum}a), which again may be due to
the contamination by post-EHB stars, as demonstrated in Fig.~\ref{cumlum}b)
where the eight potential post-EHB stars were excluded. 
%It is worth mentioning that 
In addition a bump is obvious at L/L$_{edd}$~=~-2.65, which separates sdO from
sdB stars. Hence the sdO stars can be regarded as the hot and luminous
extension of the sdB sequence.
%for which we can not
%offer an explanation yet (see Fig.~\ref{cumlum}). 
%of a bump. A clear explanation for this bump can not be given at this 
%stage, 
%but 
%We can, however, exclude the possibility of a separation of sdB from sdO stars in
%luminosity space. 
%Both sdB and sdO stars are mixed below and above this bump 
%with respect to luminosity.
%and
%sdOs due to a mixing of both types below and up this jump with respect
%to the luminosity. 
%Possibly the number of stars in our sample is too low to perfectly smooth 
%this part of the function for statistical reasons.

Simulation sets with a low CEE
efficiency (such as set 1) are reasonably consistent with the observed 
distribution in the (T$_\mathrm{eff}$, $\log{g}$) diagram as well as with 
the cumulative luminosity function if one accounts for some probably evolved 
stars.
Those 
sets, however, adopting the highest CEE efficiency (100\%) as well as those
assuming uncorrelated component masses are at variance 
with the observations and therefore
can be excluded. 

\subsubsection{Hypothesis II: sdB, sdO and He-sdO stars combined.}

The combination of He-sdO and sdO stars causes several problems when compared 
to the HPMM models.
None of the simulation sets predicts stars as hot as most of the He-sdO
stars are. Inclusion of the He-sdO stars also results in a shift of the 
cumulative luminosity function to even higher luminosities. 
None of the HPMM models predicts such luminosity functions.
As none of the simulation sets is able to reproduce the observations for
hypothesis II, we have to dismiss this assumption. 
Accordingly He-sdO stars do not belong to the same population as the sdO stars.
%The bump in the cumulative luminosity function remains, thus including the 
%HesdO stars does not 
%smooth the function either.

%\subsubsection{Hypothesis I versus II.}

%tend to not to be the same population than sdOs and
%sdBs. 

%For hypothesis I some sets do reproduce the observations well
%indicating that the sdO stars belong to the same population as the sdB stars.
%also confirmed by the comparison of the cumulative luminosity
%functions.

\subsection{Evolutionary status of He-sdO stars}

Since He-sdO stars can not be explained by HPMM models, 
other possibilities for the
origin of 
He-sdO stars have to be considered. 
Another close binary evolution model 
(the post-RGB scenario) 
as well as non-canonical evolution of single stars (the late hot flasher 
scenario) shall be discussed.

%\subsubsection{The post-RGB scenario.}

In a close binary mass transfer may occur when one of the stars fills its 
Roche lobe on the first giant branch (RGB) removing its envelope mostly. 
The remnant  
will evolve into a helium core white dwarf (see e.g. Heber et al.~2003).
We 
%compared evolutionary tracks for this post-RGB evolution (Driebe et al. 1998) 
%to the atmospheric parameters of the He-sdO sample and 
find the distribution 
of the He-sdO stars (not shown) to agree reasonably well with the 
theoretical predictions (Driebe et al.~1998). 
However, in this scenario most of the He-sdO stars should be radial 
velocity (RV) variable. 
But only one out of 28
He-sdO stars shows such a RV-variability (Napiwotzki et al.~2004).
Thus this scenario has to be regarded as 
unrealistic.

%\subsubsection{The late hot flasher scenario.}

In the {\it late hot flasher scenario} the core helium flash occurs when the 
star has already left the RGB and is approaching the white 
dwarf cooling sequence (delayed He core flash). During the flash, He 
and C is dredged-up to the surface (Sweigart, 1997).
We find that 
this scenario also can not explain the observed distribution.
%Moreover this model predicts much higher gravities than found in
%the SPY sample. 
Therefore we discard this scenario. 

Hence none of these scenarios matches the observations of He-sdO stars.

\section{Conclusion}

We conclude that those binary population synthesis models of
HPMM with a low CEE efficiency and a constant mass ratio distribution 
can explain the observed properties of the SPY sample of 
sdB and sdO stars. 
Possible explanations for the evolution of He-sdO stars are more difficult to
find. None of the considered evolution scenarios showed good agreement with
the measured parameters of our programme stars. 
%We conclude,
%that the merger scenario comes closest to explain He-sdO stars. 
The lack of radial velocity
variability of He-sdO stars (Napiwotzki et al.~2004) may support 
the merger hypothesis. However, the measured temperatures are too high. The
same holds for the gravities of some of the He-sdO stars. 
It should be noted that in the HPMM models assumptions about the remaining 
hydrogen-rich envelope had to be made. The position of a model star in the 
(T$_\mathrm{eff}$, $\log{g}$) plane, however, depends on the envelope mass. 
By reducing the adopted envelope masses in the HPMM models it may be possible 
to match the observed positions of the He-sdO stars. 
%In addition the
%evolution beyond the EHB should be incorporated in the models.
The spectra of the
He-sdO stars display lots of metal lines. Metal line blanketing, however, has
been neglected in the model atmosphere calculations, which may have led 
to an overestimate of the effective temperatures. A metal line blanketed NLTE
model grid needs to be calculated to derive more accurate parameters of
He-sdO stars.

A larger sample of
subdwarfs is needed to discriminate between the various scenarios and 
hypotheses outlined above. 
The Sloan 
Digital Sky survey (SDSS) will provide a promising source in
the near future. Kleinman et al.~(2004) already classified 240 subdwarfs
from the first data release. 
At the end of the SDSS project spectra of more than 1000 hot subdwarfs 
will become available for spectral analyses.

\acknowledgements{A.S. thanks the workshop organizers for financial support.
We are grateful to I. Traulsen and T. Rauch for their help with the 
NLTE model atmosphere code.}

\end{document}